\documentclass[10pt,conference]{IEEEtran}
\usepackage{times,amsmath,amssymb,graphicx,epsfig,cite,amsthm}
\usepackage{subfigure,url}

\DeclareMathOperator*{\argmax}{arg\,max}

\DeclareMathOperator*{\newrightarrow}{\longrightarrow}

\newtheorem{theorem}{Theorem}

\renewcommand{\baselinestretch}{0.98}

\begin{document}

\title{Achievable Rate and  Optimal Physical Layer
  Rate Allocation in Interference-Free Wireless  Networks}
\author{\authorblockN{Tao Cui and Tracey Ho}
  \authorblockA{Department of Electrical Engineering\\
    California Institute of Technology\\
    Pasadena, CA 91125, USA\\
    Email: \{taocui, tho\}@caltech.edu}\and\authorblockN{J{\"o}rg
    Kliewer}
  \authorblockA{Klipsch School of Electrical and Computer Engineering \\
    New Mexico State University\\
    Las Cruces, NM 88003, USA\\
    Email: jkliewer@nmsu.edu}}
\IEEEoverridecommandlockouts
\IEEEaftertitletext{\vspace{-2.1\baselineskip}}
 \maketitle

\begin{abstract}
  We analyze the achievable rate in interference-free wireless
  networks with physical layer fading channels and orthogonal multiple
  access. As a starting point, the point-to-point channel is
  considered.  We find the optimal physical and network layer rate
  trade-off which maximizes the achievable overall rate for both a
  fixed rate transmission scheme and an improved scheme based on
  multiple virtual users and superposition coding. These initial
  results are extended to the network setting, where, based on a
  cut-set formulation, the achievable rate at each node and its upper
  bound are derived. We propose a distributed optimization algorithm
  which allows to jointly determine the maximum achievable rate, the
  optimal physical layer rates on each network link, and an
  opportunistic back-pressure-type routing strategy on the network
  layer.  This inherently justifies the layered architecture in
  existing wireless networks.  Finally, we show that the proposed
  layered optimization approach can achieve almost all of the ergodic
  network capacity in high SNR.

\end{abstract}

\vspace{-0.5ex}
\section{Introduction}\vspace{-0.5ex}
In wireless communication networks, channel impairments such as
fading, shadowing and path loss limit the capacity. Determining the
capacity region and corresponding achievable strategy for general
multiterminal networks is a long-standing open problem. An outer bound
for the capacity region is known to have cut-set interpretation, but
this cut-set bound is not always achievable. For example, there is no
known scheme to achieve this outer bound for the simple relay channel.
In contrast, in wireline networks with orthogonal channels, the
cut-set bound is achievable even in the multicast scenario, where
multiple destinations demand the same information from several
sources. In this case, the capacity of wireline networks can be
achieved via a separation of physical layer channel coding and higher
layer network coding. Furthermore, in \cite{dana06} it is shown that
the cut-set bound is tight in wireless erasure networks. Also, related
work in \cite{RK06} shows that the capacity of networks with
deterministic channels and broadcasting has a cut-set formulation. The
optimal rate trade-off between physical and network layer is found
analytically in~\cite{VM05} for a point-to-point channel under an
overall delay constraint, and by simulation
in~\cite{LGSW07}\footnote{We thank J.~Christopher Ramming for bringing
  this paper to our attention.} for minimum energy consumption in
cellular broadcast networks.  However, similar achievability results
do not yet exist for more general wireless networks with fading
channels.

In this paper, we study the achievable rate and optimal physical layer
rate allocation for a special class of wireless networks, where for
each node there is no interference between the received signals from
its neighbors.  This could be realized by using orthogonal access
strategies such as \mbox{time-,} \mbox{frequency-,} or code-division
multiple access. Further, the channels between each node and one of
its neighbors are modeled as slow fading channels.  For the case where
channel state information (CSI) is only known at the receiver, each
node simply transmits at a constant rate regardless of the fading
state. Therefore, each node's transmitted information gets lost when
the transmission rate is high so that the current channel cannot
support it, which results in outage.  For the point-to-point case we
present a throughput-efficient coding scheme by partitioning the
transmit power across multiple virtual users with different rates. The
corresponding information is then transmitted using superposition
coding. This approach outperforms the constant rate scheme in terms of
achievable rate. For an interference-free network with fading channels
we consider the outage probability as the link erasure probability in
\cite{dana06}.  Different from \cite{dana06} where the erasure
probability is given, the outage probability depends on the
transmission rate at each node.  We first obtain the achievable rate
of this network and its outer bound based on a cut-set formulation.
Further, by using a flow formulation, we derive a distributed
optimization algorithm based on a modified dual decomposition approach
to maximize the achievable rate.  As a byproduct, the algorithm also
returns the optimal physical layer rate at each node, and an optimal
opportunistic back-pressure-type routing strategy on the network
layer. Finally, we bound the gap between the achievable rate and the
outer bound. The result reveals that the proposed layered structure
can achieve almost all of the ergodic network capacity in high SNR,
which justifies the pervasive layered architecture in existing
wireless networks.

\vspace{-0.5ex}
\section{Preliminaries}\label{sect2}
\vspace{-0.5ex}

\subsection{Network Model}\vspace{-0.5ex}
We model the wireless network by a directed acyclic graph
$\mathcal{G}=(\mathcal{V},\mathcal{E})$, where $\mathcal{V}$ is the
vertex set and $\mathcal{E}$ is a directed edge set. For each node
$i\in \mathcal{V}$, $\mathcal{N}_{\mathcal{O}}(i)$ and
$\mathcal{N}_{\mathcal{I}}(i)$ denote the set of in-neighbors and
out-neighbors of $i$, i.e., $
\mathcal{N}_{\mathcal{I}}(i)=\left\{j|(j,i)\in \mathcal{E}\right\}$,
$ \mathcal{N}_{\mathcal{O}}(i)=\left\{j|(i,j)\in
\mathcal{E}\right\}$. A cut for $x,y\in \mathcal{V}$ is a partition
of $ \mathcal{V}$ into two sets $ \mathcal{V}_x$ and $
\mathcal{V}_y= \mathcal{V}_x^c$ such that $x\in \mathcal{V}_x$ and
$y\in \mathcal{V}_y$. Furthermore, we define $ \mathcal{V}_x^*$ to
be the set of nodes in $\mathcal{V}_x$ that have at least one
outgoing edge in $\mathcal{V}_x^c$, i.e., $
\mathcal{V}_x^*=\{i|\exists j,\ \text{such that } (i,j)\in
\mathcal{E}\}$. We focus on acyclic graphs in the following.

In this paper, each edge $(i,j)\in \mathcal{E}$ represents a
memoryless Gaussian channel from node $i$ to node $j$.  Also, we
assume that each node has a unit bandwidth.  Let $x_i$ be the
transmitted signal by node $i$ and $y_{i,j}$ be the received signal at
node $j$ from node $i$, where the average power of $x_i$ is $P_i$,
e.g., $E\left\{|x_i|^2\right\}=P_i$. We have the channel model

$ $\vspace{-2mm}
\begin{small}
\begin{equation}\label{eq201}
y_{i,j}=\sqrt{h}_{i,j}x_j+v_{i,j},\vspace{-2mm}
\end{equation}
\end{small}
where $\sqrt{h}_{i,j}$ is the channel fading between node $i$ and
node $j$ and $v_{i,j}$ is the additive white Gaussian noise at node
$j$ for the signal reception from node $i$. Without loss of
generality, we assume that $v_{i,j}$ has zero mean and unit
variance. The variance $\sigma_{i,j}^2$ of $h_{i,j}$ is given and
normalized such that $0\leq\sigma_{i,j}^2\leq 1$. The channel gain
$\sqrt{h}_{i,j}$ is assumed to be constant over each packet of size
$n$ and to vary identically and independently
 between different packets.
\hspace{-0.1mm}We \hspace{-0.1mm}consider\hspace{-0.1mm}
a\hspace{-0.1mm} Rayleigh\hspace{-0.1mm} fading\hspace{-0.1mm}
channel\hspace{-0.1mm}, \hspace{-0.2mm}where \hspace{-0.2mm}the
\hspace{-0.2mm} probability \hspace{-0.2mm}density
\hspace{-0.2mm}function \hspace{-0.2mm}(pdf) \hspace{-0.2mm}of
\hspace{-0.2mm}$h_{i,j}$ \hspace{-0.2mm}is

\vspace{-2mm}
\begin{small}
\begin{equation}\label{eq202}
f(h_{i,j})=\frac{1}{\sigma_{i,j}^2}e^{-\frac{h_{i,j}}{\sigma_{i,j}^2}},\vspace{-2mm}
\end{equation}
\end{small}%
and $h_{i,j}$ and $h_{i,j^{\prime}}$ are independent $\forall j,
j^{\prime}\in\mathcal{N}_{\mathcal{O}}(i),\,j\neq j^{\prime}$.
However, the results extend to arbitrary fading distributions and
correlated fading channels in a straightforward way.

In the following we address multicast problems where a set of sinks
$\mathcal{D}=\{d_1,\ldots,d_{|\mathcal{D}|}\}\subset\mathcal{V}$
demands all of the information from a set of sources. In this paper,
we focus on the case of a single source
$s\in\mathcal{V}$. The case of multiple sources can be generalized
in the same way as in  \cite{dana06}.

Our encoding scheme is given by layered scheme based on a product code
\cite{lin04}, where a horizontal channel code is used for hop-by-hop
physical layer error correction and a vertical erasure correcting code
is used for end-to-end error correction on the network layer. This
provides an implicit block interleaving as an outage for a physical
layer block leads to individual symbol erasures on the network layer.

\section{Achievable Rate in Point-to-Point Channels}\label{sect3}

We first study the point-to-point channel, where the system model
$y=\sqrt{h}x+v$ is as described in the previous section but with the
subscripts $i,j$ removed for brevity.  Our objective is to maximize
the average throughput by optimizing the rate allocation for both
network and physical layer.

When CSI is only known at the receiver the transmitter cannot adapt
its transmission rate according to $h$.  In this case the capacity is
\cite{goldsmith97}

\vspace{-2mm}
\begin{small}
\begin{equation}\label{eq302_1}
C=E_h\left\{\frac{1}{2}\log\left(1+hP\right)\right\},\vspace{-1.5mm}
\end{equation}
\end{small}
if maximum likelihood decoding is used \cite{goldsmith97}. Simple
encoding and decoding techniques for AWGN channels cannot be applied
directly. Without transmitter CSI, one method is to let the
transmitter transmit at a constant rate $R$. Therefore, if $ R\leq
C(h)=\frac{1}{2}\log\left(1+hP\right)$, the receiver can successfully
decode the transmitted signal. To maximize the average throughput, we
need to solve

\vspace{-4mm}
\begin{small}
\begin{equation}\label{eq305}
\max_R R\int_{\frac{2^{2R}-1}{P}}^{+\infty}f(h)dh=\max_R
Re^{-\frac{2^{2R}-1}{P\sigma^2}}=\max_R F(R), \vspace{-1.5mm}
\end{equation}
\end{small}
where we have used the Rayleigh fading pdf in \eqref{eq202}. In
\eqref{eq305}, $F(R)$ denotes the overall rate, $R$ the rate on the
physical layer, and the term $\exp(-\frac{2^{2R}-1}{P\sigma^2})$
corresponds to the network layer rate as a function of $R$. By taking
the derivative of $F(R)$ with respect to $R$ and setting the result to
zero, we obtain the optimal physical layer rate as

\vspace{-2mm}
\begin{small}
\begin{equation}\label{eq308}
R^*=\frac{1}{2\ln 2} W(P\sigma^2), \vspace{-1.5mm}
\end{equation}
\end{small}%
where $W(\cdot)$ is the Lambert W function, the inverse function of
$g(w) = we^w$. By substituting $R^*$ into $F(R)$ in \eqref{eq305}, we
obtain the maximal average overall throughput as

\vspace{-4mm}
\begin{small}
\begin{equation}\label{eq308_15}
F\left(R^*\right)=R^*e^{-\frac{2^{2R^*}-1}{P\sigma^2}}=\frac{1}{2\ln
2} W(P\sigma^2)e^{-\frac{1}{ W(P\sigma^2)}}e^{\frac{1}{P\sigma^2}}.
\vspace{-1mm}
\end{equation}
\end{small}

We now discuss a coding scheme which is able to improve the throughput
compared to the constant rate case. The idea was first presented in
\cite{Sha97} and is based on superposition coding by introducing
an infinite number of  virtual users at the transmitter. We then
propose a modification of this approach by considering the practically more relevant case
for a finite number of virtual users.

Suppose that each of the users transmits with a power $dz$ on the
physical layer. At the decoder we can employ successive interference
cancellation to decode the virtual users' messages, where the virtual
user with a higher interference level $z$, $0\leq z\leq P$, is decoded
first by treating all the virtual users below as interference.  Thus,
different virtual users incur different interference levels $z$ at the
receiver. In particular, the virtual user at interference level $z$
transmits at rate $r(z)dz$, where $r(z)$ denotes the marginal rate.
Note that at channel fading state $h$, a virtual user with power $dz$
and Gaussian interference $1+hz$ can achieve the rate

\vspace{-2mm}
\begin{small}
\begin{equation}\label{eq308_12}
\frac{1}{2}\log\left(1+\frac{hdz}{1+hz}\right)\newrightarrow^{dz\rightarrow
0} \frac{1}{2\ln 2}\frac{h}{1+hz}dz. \vspace{-1.5mm}
\end{equation}
\end{small}
Therefore, the message sent by the virtual user at interference
level $z$ can be decoded if $\frac{1}{2\ln 2}\frac{h}{1+hz}\geq
r(z)$, and all the virtual users with higher interference levels are
decoded successfully, if $\frac{1}{2\ln
2}\frac{h}{1+hz^\prime}\geq r(z^\prime)$, $\forall z^\prime\geq z$,
which is denoted as Condition A in the following. If $\frac{1}{2\ln
2}\frac{h}{1+hz}= r(z)$ has a unique solution for $z$ between $0$ and $P$
and $r(z)$ is a strictly decreasing function, Condition A is satisfied.
Then, the probability that the virtual user at interference level $z$
can be decoded is given as

\vspace{-2mm}
\begin{small}
\begin{equation}\label{eq308_2}
\mathcal{P}_d(r(z),z)=Pr\left(\frac{1}{2\ln 2}\frac{h}{1+hz}\geq r(z)\right).
\vspace{-1.5mm}
\end{equation}
\end{small}
Note that in this case the network layer rate is obtained by
integrating \eqref{eq308_2} over all $z$, $0\leq z\leq P$. Thus,
the achievable average overall throughput for Rayleigh fading is

\mbox{}
\vspace{-2ex}
\begin{small}\begin{equation}\label{eq308_3}
F(r(\cdot))=\int_0^P\mathcal{P}_d(r(z),z)r(z)dz=\int_0^Pe^{-\frac{r(z)}{(1-r(z)z)\sigma^2}}r(z)dz.
\vspace{-1.5mm}
\end{equation}
\end{small}%
We need to maximize (\ref{eq308_3}) over all possible marginal rate
functions $r(\cdot)$ that satisfy Condition A. Clearly, the marginal
rate function $r(z)=\frac{1}{2\ln 2\left(\frac{P}{2^{2R}-1}+z\right)}$
satisfies Condition A, in which case (\ref{eq308_3}) becomes the
single fixed rate problem (\ref{eq305}). This means that the
superposition approach contains the fixed rate scheme as a special case.

By maximizing (\ref{eq308_3}) for each $z$ individually, we obtain the
optimal marginal rate function as

\vspace{-2mm}
\begin{small}\begin{equation}\label{eq308_5}
r^*(z)=\frac{1}{(2\ln 2)\sigma^2
z^2}\left(1+2\sigma^2z-\sqrt{1+4\sigma^2 z}\right). \vspace{-2mm}
\end{equation}
\end{small}
We can see that $\frac{1}{2\ln 2}\frac{h}{1+hz}= r^*(z)$ has a
unique solution $z=\frac{\sigma^2-h}{h^2}$ and that $r^*(z)$ satisfies
Condition A. Therefore, $r^*(z)$ is the optimal solution. By
integrating $r^*(z)$ over all $z$,  $0\leq z\leq P$, we get the
optimal physical layer rate.

We now propose a modification of the above scheme for a finite number
of virtual users.  For example, for two virtual users we choose the
marginal rate function as

\vspace{-2mm}
\begin{small}
\begin{equation}\label{eq308_6}
r(z)=\left\{
       \begin{array}{cc}
         \frac{1}{2\ln 2\left(\frac{P}{2^{2R_1}-1}+z\right)}, & \text{if } 0\leq z\leq \alpha P, \\
         \frac{1}{2\ln 2\left(\frac{P}{2^{2R_2}-1}+z\right)}, & \text{if }\alpha P\leq z\leq P. \\
       \end{array}
     \right.\vspace{-1mm}
\end{equation}
\end{small}
To satisfy Condition A, we must have $R_2\leq R_1$.  Substituting
(\ref{eq308_6}) into (\ref{eq308_3}), from \eqref{eq308_5} we obtain
the physical layer rate

\vspace{-3.5mm}
\begin{small}\begin{equation}\label{eq308_7}
\begin{split}
F(R_1,R_2,\alpha)
=&\frac{1}{2}\log\left(1+\alpha(2^{2R_1}-1)\right)e^{-\frac{2^{2R_1}-1}{P\sigma}}\\
&\hspace{8mm}+\frac{1}{2}\log\left(1+\frac{(1-\alpha)}{\frac{1}{2^{2R_2}-1}+\alpha
}\right)e^{-\frac{2^{2R_2}-1}{P\sigma}}.
\end{split}
\end{equation}
\end{small}%
\vspace{-2mm}

\noindent Fixing $\alpha$, we can find the optimal $R_1(\alpha)$
and $R_2(\alpha)$ by maximizing the two summands in
(\ref{eq308_7}) separately. Substituting $R_1(\alpha)$ and
$R_2(\alpha)$ back into (\ref{eq308_7}), we can find the optimal
$\alpha$ by a line search. We show experimentally in Section V
that the performance for two virtual users is not far from the
infinite user bound. The approach extends to the case with more
virtual users in a straightforward way.

\vspace*{-0.4ex}
\section{Achievable Rate in Wireless Networks}\label{sect4}
\vspace*{-0.8ex} In this section, we consider general wireless
networks without interference. As above, we assume that CSI is
only available at the receiver. \vspace*{-0.5ex}
\subsection{Achievable Rate Region and Upper Bound}\label{sect41}
\vspace*{-0.9ex} For the sake of brevity we only consider an extension
of the fixed rate case in Section~\ref{sect3}; results similar to
those presented below can also be obtained for the virtual user-based
scheme. Each node $i$ simply broadcasts to its neighbors at a common
rate $R_i$ due to the lack of CSI at the transmitter. The probability
that node $j\in \mathcal{N}_{\mathcal{O}}(i)$ can receive the packets
sent by node $i$ is given as

\vspace{-2.5mm}
\begin{small}\begin{equation}\label{eq402}
\mathcal{P}_{i,j}(R_i)=\int_{\frac{2^{2R_i}-1}{P_i}}^{+\infty}f(h_{i,j})dh_{i,j}=e^{-\frac{2^{2R_i}-1}{P_i\sigma_{i,j}^2}},
\vspace{-1mm}
\end{equation}
\end{small}
where we have used the Rayleigh pdf from \eqref{eq202}. Due to the
ergodicity of the channel, we can associate each network layer packet
transmitted by node $i$ with an erasure probability
$\epsilon_{i,j}(R_i)=1-\mathcal{P}_{i,j}(R_i)$ at node $j$. Therefore,
at the network layer, we see an erasure network, which is similar to
that in \cite{dana06}. It can be readily verified that the results in
\cite{dana06}\hspace{-0.1mm} can \hspace{-0.1mm}be
\hspace{-0.1mm}extended\hspace{-0.1mm} to \hspace{-0.1mm}our
\hspace{-0.1mm} case\hspace{-0.1mm} with \hspace{-0.1mm}only
\hspace{-0.1mm} a\hspace{-0.1mm} slight \hspace{-0.1mm}modification
\hspace{-0.1mm}in the proof, which leads to the following Theorem.

\vspace*{-0.5ex}
\begin{theorem}
The capacity of the wireless network with fixed rate
transmission and Rayleigh fading on each link is given by the
minimum value of the cuts between the source node and any of the
destinations, i.e.,

\vspace{-4.5mm}
\begin{small}
\begin{align} C=&\max_{\{f_i(\cdot)|i\in
\mathcal{V}\}}\min_{\{\mathcal{V}_s|\mathcal{V}_s\subset
\mathcal{V},\,  s\in \mathcal{V}_s \atop\text{ and }
\mathcal{D}\cap\mathcal{V}^c_s\neq \emptyset\}}\sum_{i\in
\mathcal{V}^*_s}\int_0^{+\infty}
f_i(R_i)R_i\nonumber\\
&\hspace{30mm}\times\left(1-\prod_{\{j|j\in \mathcal{V}^c
\cap\mathcal{N}_{\mathcal{O}}(i) \}}\epsilon_{i,j}(R_i)\right)dR_i \nonumber\\[-0.5ex]
&\text{subject to\  }\int_0^{+\infty} f_i(R_i)dR_i=1,\label{eq403}
\end{align}
\end{small}
\end{theorem}
In Theorem 1, the function $f_i(R_i)$ plays the role of time sharing
between different rates. Since $R_i\left(1-\prod_{\{j|j\in \mathcal{V}^c
\cap\mathcal{N}_{\mathcal{O}}(i) \}}\epsilon_{i,j}(R_i)\right)$ is
generally not a convex function in $R_i$, time sharing may
achieve some rate that is not achievable without using time sharing.
Note that a similar result as in Theorem~1 can also be obtained for the virtual-user
based scheme from Section~\ref{sect3}.

Since Theorem 1 only addresses the fixed rate transmission scheme for
the receiver-only CSI case, it is instructive to also consider the
upper bound on the achievable rate for any transmission scheme.
Define $\underline{Y}_j=\{Y_{i,j}|{i\in
  \mathcal{N}_{\mathcal{I}}(j)}\}$. By using the cut-set bound from
\cite[Theorem 14.10.1]{cover91} we obtain

\vspace{-4mm}
\begin{small}\begin{equation}\label{eq407}
R_s\leq
I\left(X^{\mathcal{V}_s};\underline{Y}^{\mathcal{V}_s^c},H^{[\mathcal{V}_s,\mathcal{V}_s^c]}|X^{\mathcal{V}^c_s}\right),\,\forall
\mathcal{V}_s\subset \mathcal{V},\, s\in \mathcal{V}_s \text{ and }
\mathcal{D}\cap\mathcal{V}^c_s\neq \emptyset,
\end{equation}
\end{small}

\vspace{-2ex}
\noindent
where $X^{\mathcal{V}_s}=\{X_i|i\in \mathcal{V}_s\}$,
$Y^{\mathcal{V}_s}=\{\underline{Y}_j|j\in \mathcal{V}^c_s\}$,
$H^{[\mathcal{V}_s,\mathcal{V}_s^c]}=\{H_{i,j}|i\in
\mathcal{V}_s,\,j\in\mathcal{V}_s^c,\,(i,j)\in \mathcal{E}\}$, and
$H_{i,j}$ denotes the corresponding random variable for $h_{i,j}$.
As CSI is known at each receiver, we can interpret the CSI as an
additional received quantity at the receiver. Since we assume that
different nodes have orthogonal channels and that the network is
acyclic, an upper bound on the network capacity of receiver CSI only
can be obtained from \eqref{eq407} as

\vspace{-4mm}
\begin{small}\begin{equation}\label{eq410}
C\leq \min_{\{\mathcal{V}_s|\mathcal{V}_s\subset \mathcal{V},\, s\in
\mathcal{V}_s \atop \text{ and } \mathcal{D}\cap\mathcal{V}_s^c\neq
\emptyset\}} \sum_{i\in \mathcal{V}_s^*}
E_h\left\{\frac{1}{2}\log\left(1+P_i\hspace{-1mm}\sum_{j\in
\mathcal{V}_s^c
\cap\mathcal{N}_{\mathcal{O}}(i)}\hspace{-2mm}h_{i,j}\right)\hspace{-1mm}\right\}\vspace{-1mm}
\end{equation}
\end{small}%

Note that the upper bound is difficult to achieve in a general network because the
receivers in the cut-set are assumed to have full cooperation to
obtain the cut-set rate. In particular, this holds for an achievable
scheme. However, in Section \ref{sect44} we will
show that the achievable rate is not far away from this upper bound
in high SNR.


\subsection{Opportunistic Routing and Distributed
Algorithm}\label{sect43}

To solve (\ref{eq403}) directly, we need to consider all the cuts
which may be exponential in the number of nodes in the network. As
an alternative, we consider a flow-based formulation where the
constraints are only polynomial in $|\mathcal{E}|$ and
$|\mathcal{V}|$. At each node $i$ in the network let the rate on
each outgoing link be denoted as $x^d_{i,j}$ for each destination
node $d\in \mathcal{D}$. Then, in order to find the maximum
multicast rate, we need to solve

\vspace{-4mm}
\begin{small}
\begin{align}
\max_{C,\mathbf{x},f}\text{ }& C \nonumber\\[-2ex]
\text{s.t. }& \sum_{\{j|j\in
\mathcal{N}_{\mathcal{O}}(i)\}}x_{i,j}^d-\sum_{\{j|j\in
\mathcal{N}_{\mathcal{I}}(i)\}}x_{j,i}^d=\left\{
                                           \begin{array}{cc}
                                             C, & \text{if } i=s, \\
                                             -C, & \text{if } i=d, \\
                                             0, & \text{otherwise}, \\
                                           \end{array}
                                         \right.\nonumber\\
&\sum_{j\in \mathcal{Z}}x^d_{i,j}\leq \int_0^{+\infty}
f_i(R_i)R_i\left(1-\prod_{j\in
\mathcal{Z}}\epsilon_{i,j}(R_i)\right)dR_i,\, \forall
\mathcal{Z}\subseteq \mathcal{N}_{\mathcal{O}}(i),\nonumber\\
&\lefteqn{\int_0^{+\infty} f_i(R_i)dR_i=1,}\label{eq435}
\end{align}
\end{small}%
\vspace{-4mm}

\noindent where $x_{i,j}^d$ is the flow rate on link $(i,j)$ for
destination $d\in \mathcal{D}$ and the vector $\mathbf{x}$ contains
the flow rates for all $d$ and for all edges in the network. The
right hand side of the last constraint in \eqref{eq435} represents
the total amount of information that can be decoded by at least one
node in $\mathcal{Z}$. It can be readily verified that (\ref{eq403})
and (\ref{eq435}) have the same optimal value.

Note that (\ref{eq435}) is similar to the problem in \cite{chen06},
where a dual decomposition based cross layer optimization is proposed.
Our problem differs from that in \cite{chen06} in the realization of
the physical layer rate, which also results in a different routing
scheme compared to \cite{chen06} as shown in the following. By
adopting a similar dual decomposition approach as in \cite{chen06}, we
can write the Lagrange dual of (\ref{eq435}) as

\mbox{} \vspace*{-2.5ex}
\begin{small}
\begin{align}\label{eq436}
D(\mathbf{q})=&\max_{\mathbf{x},f,C} \text{ }\left(1-\sum_{d\in
\mathcal{D}}q_s^d\right)C+\sum_{(i,j)\in \mathcal{E}}\sum_{d\in
\mathcal{D}}\left(q_i^d-q_j^d\right)x_{i,j}^d\nonumber\\
&\hspace{-5mm}\text{s.t. }\sum_{j\in \mathcal{Z}}x^d_{i,j}\leq
\int_0^{+\infty} f_i(R_i)R_i\left(1-\prod_{j\in
\mathcal{Z}}\epsilon_{i,j}(R_i)\right)dR_i,\, \forall
\mathcal{Z}\subseteq \mathcal{N}_{\mathcal{O}}(i),\nonumber\\
&\hspace{0mm}\lefteqn{\int_0^{+\infty} f_i(R_i)dR_i=1,}
\end{align}
\end{small}
\vspace{-3.5mm}

\noindent where $\chi_i^d$ is defined as the right hand side of the
first constraint in (\ref{eq435}), and $q_i^d$ denotes the
Lagrangian multiplier at node $i$ for destination $d$. The
collection of all $q_i^d$ is given by the vector $\mathbf{q}$.  Due
to the convexity of (\ref{eq435}) there is no duality gap.

To find $D(\mathbf{q})$ in \eqref{eq436}, we find that the
maximization problem separates in two individual maximizations over
$\mathbf{x}$  and $f$, resp., and $C$ for given $\mathbf{q}$. In order to update
$C$ we use a primal subgradient method.  Let $C^t$ be the value of
$C$ at the $t$-th iteration. We update $C$ using

\vspace{-3.5mm}
\begin{small}
\begin{equation}\label{eq4312}
C^{t+1}=\left[C^t+\gamma^t \left(1-\sum_{d\in
\mathcal{D}}q_s^d\right)\right]^+, \vspace{-2.5mm}
\end{equation}
\end{small}

\noindent where $[\cdot]^+$ denotes the mapping to non-negative
numbers and $\gamma^t>0$ is a stepsize. For the second maximization
over $\mathbf{x}$, let $\pi^d$ be a permutation of $j\in
\mathcal{N}_{\mathcal{O}}(i)$ such that
$w^d_{i,j}=\left[q_i^d-q_j^d\right]^+$ satisfies $w_{i,\pi^d_1}\geq
w_{i,\pi^d_2}\geq\cdots\geq
w_{i,\pi^d_{\left|\mathcal{N}_{\mathcal{O}}(i)\right|}}$. Due to the
polymatroid structure of the constraint in (\ref{eq436}),
\eqref{eq436} is maximized when

\vspace{-4mm}\begin{small}%
\begin{equation}\label{eq4314}
\begin{split}
x_{i,\pi^d_1}^d=&\int_0^{+\infty} \hspace{-0mm}f_i(R_i)R_i\left(1-\epsilon_{i,\pi^d_1}(R_i)\right)dR_i,\\
x_{i,\pi^d_k}^d=&\int_0^{+\infty}\hspace{-0mm}
f_i(R_i)R_i\prod_{j=1}^{k-1}\epsilon_{i,\pi^d_j}(R_i)\left(1-\epsilon_{i,\pi^d_k}
  (R_i)\right)dR_i,\\
  &\hspace{45mm}k = 2,\dots,\left|\mathcal{N}_{\mathcal{O}}(i)\right|.
\end{split}
\end{equation}
\end{small}
\vspace{-2mm}

\noindent We can see that the solution (\ref{eq4314}) corresponds to
prioritizing the neighbors of node $i$ for destination $d$ according
to $w^d_{i,j}$.  Thus, if a higher priority neighbor has received the
message from node $i$ for destination $d$, all the lower priority
neighbors will drop the same message even though they also have
received it. We can interpret $q_i^d$ as the queue length of the flow
for destination $d$ at node $i$ and $w^d_{i,j}$ as the aggregate queue
length difference between $i$ and $j$ for destination $d$,
respectively.  Different from opportunistic routing \cite{biswas04}
where neighbors are prioritized heuristically, we optimize the
prioritization by taking into account the queue information at each
node. Note that each node's queue length is affected by both the
incoming traffic rate and the depletion rate. The depletion rate
implicitly counts the optimization over the network beyond node $i$.
Thus, this prioritization approach can be seen as special variant of a
back-pressure-based routing scheme \cite{ho05}.

Substituting (\ref{eq4314}) into (\ref{eq436}), we obtain an
optimization problem which depends only on $f_i(\cdot)$. The solution
is $ f_i(R_i)=\delta(R_i-R_i^*)$,  where $\delta(\cdot)$ is the
Dirac delta function, leading to

\vspace*{-4mm}\begin{small}
\begin{equation}\label{eq4317_2}
R_i^*=\argmax_{R_i}
 R_i\hspace{-1mm}\sum_{j=1}^{\left|\mathcal{N}_{\mathcal{O}}(i)\right|}\hspace{-1mm}\sum_{d\in
\mathcal{D}}\left(\hspace{-0.5mm}w^{d}_{i,\pi^d_j}-w^{d}_{i,\pi^d_{j+1}}\hspace{-0.5mm}\right)\hspace{-1mm}\left(\hspace{-1mm}1-\hspace{-0.5mm}\prod_{k=1}^{j}\epsilon_{i,\pi^d_k}(R_i)\hspace{-1mm}\right).\hspace{-1mm}
\end{equation}
\end{small}
\vspace{-4mm}

\noindent As (\ref{eq4317_2}) is not convex, we can solve it using a
line search. Note that at different iterations $t$ the chosen rate
$R_i^*$ in (\ref{eq4317_2}) can be different due to oscillations of
$\mathbf{q}$. Time sharing is then realized by averaging $R_i^*$ over
all iterations.

After solving (\ref{eq436}), we obtain the dual function
$D(\mathbf{q})$, which needs to be minimized. Instead of minimizing
$D(\mathbf{q})$ directly, we use the dual subgradient method to update
$\mathbf{q}$. Let $q_{i}^{d,t}$ be the dual variable at the $t$-th
iteration. We update $q_{i}^{d,t}$ via

\mbox{} \vspace{-4mm}
\begin{footnotesize}
\begin{equation}\label{eq4315}
q_{i}^{d,t+1}=\left\{\hspace{-2mm}
                \begin{array}{l}
                  \left[q_{i}^{d,t}\hspace{-0.2mm}-\hspace{-0.2mm}\eta^t\left(\sum_{\{j|j\in
\mathcal{N}_{\mathcal{O}}(i)\}}x^d_{i,j}-\sum_{\{j|j\in
\mathcal{N}_{\mathcal{I}}(i)\}}x^d_{j,i}-C^t\right)\right]^+\hspace{-2mm},\\
\hspace{62.5mm}\text{if }i=s, \\
                  0,\hspace{60mm}\text{if }i=d, \\
                  \left[q_{i}^{d,t}\hspace{-0.2mm}-\hspace{-0.2mm}\eta^t\left(\sum_{\{j|j\in
\mathcal{N}_{\mathcal{O}}(i)\}}x^d_{i,j}-\sum_{\{j|j\in
\mathcal{N}_{\mathcal{I}}(i)\}}x^d_{j,i}\right)\right]^+,\\
\hspace{60mm}\text{otherwise}, \\
                \end{array}
              \right.
\end{equation}
\end{footnotesize}

\vspace{-2ex}
\noindent
where $\eta^t>0$ is a constant stepsize, and  $C^t$ and
$x_{i,j}^d$ are obtained from \eqref{eq4312} and \eqref{eq4314},
respectively. The updated $q_{i}^{d,t+1}$ is then used in
\eqref{eq436} again for the next iteration. By the subgradient
optimization theory and the convexity of the problem, the proposed
primal-dual subgradient algorithm in (\ref{eq4312})-\eqref{eq4317_2}
converges to the optimal solution in (\ref{eq436}) when  stepsizes
$\gamma^t$ and $\eta^t$ go to zero as $t\rightarrow \infty$ and
converges to a small neighborhood of the optimal solution with fixed
stepsizes \cite{chen06}.

An important benefit of the presented algorithm is that
(\ref{eq4312}), (\ref{eq4314}), and \eqref{eq4317_2} can all be solved
locally at each node $i$ given the queue length information of its
neighbors, which only involves local information exchange.  This
algorithm can not only be used to determine the maximal multicast rate
and the corresponding physical layer rate on each link but also
suggests the optimal routing protocol to achieve this rate. We can see
that a layered structure can be used, where channel coding is applied
at the physical layer and opportunistic routing and erasure correction
coding are applied at the networking layer.

\vspace{-0.4ex}
\subsection{Asymptotic Behavior}\label{sect44}
In \cite{CHK09} we prove the following theorem which quantifies
the capacity gap between the
achievable scheme proposed in Section~\ref{sect43} and the upper bound
on the achievable rate from  Section~\ref{sect41}.

\vspace{-1.2ex}
\begin{theorem}
Let
$C^\dag(\mathbf{P})=C^\dag([P_1,P_2,\dots,P_{|\mathcal{V}|}])$ be the
capacity region of the fixed size wireless network without
interference and $C(\mathbf{P})$ be the achievable rate with fixed rate
transmission at each node. Further, let $\varsigma=\max_i\max_{j\neq
j^\prime}\sigma^2_{i,j}/\sigma^2_{i,j^\prime}$. Then,

\vspace{-4mm}
\begin{small}
\begin{equation}\label{eq449}
\begin{split}
\lim_{\mathbf{P}\rightarrow\infty}C^\dag(\mathbf{P})-C(\mathbf{P})\leq &
\frac{1}{2}|\mathcal{D}||\mathcal{V}|\log
\left(|\mathcal{V}|\varsigma\right)+0.7588|\mathcal{D}||\mathcal{V}|\\[-0.5ex]
&\hspace{15mm}+o\left(|\mathcal{D}|\sum_{i\in \mathcal{V}}\ln\ln(
P_i)\right),
\end{split}
\end{equation}
\end{small}%
where the limit is taken for each entry of $\mathbf{P}$
individually. Moreover, $\lim_{\mathbf{P}\rightarrow\infty}C^\dag(\mathbf{P})/C(\mathbf{P})=1$.
\end{theorem}

Theorem 2 indicates that the fixed rate transmission scheme can
achieve almost all of the ergodic network capacity asymptotically in
high SNR. Hence, in high SNR the proposed layered structure is close
to optimal which justifies the layered structure in existing wireless
network protocols. However, different from existing protocols, in the
proposed scheme the queue length information determines not only the
opportunistic routing but also the physical layer data transmission
rate at each node.

\vspace*{-1.4ex}
\section{Simulation Results}\label{sect5}\vspace*{-0.5ex}
We assume in the following that all the nodes have the same transmit power $P$ and that
the channel gains have the same unit variance.

\begin{figure}\centering
  \includegraphics[scale=0.45]{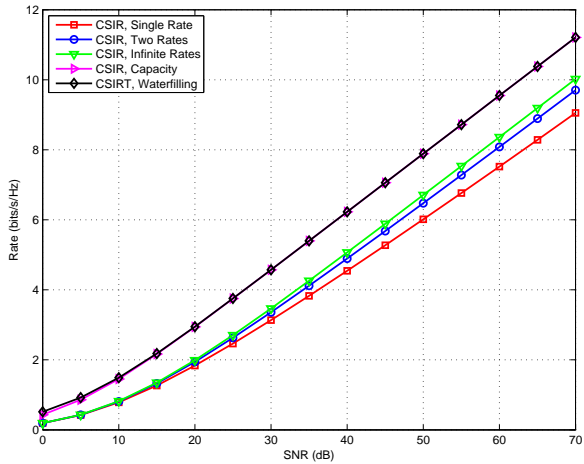}\\
  \vspace{-3mm}
  \caption{Comparison of achievable rates for different schemes in point-to-point Rayleigh fading channels.}\vspace{-6mm}\label{f3}
\end{figure}

Fig.~\ref{f3} compares the achievable rate of different schemes for a
point-to-point Rayleigh fading channel. The fixed (single) rate
transmission without transmitter CSI (\ref{eq308}) is denoted as
``CSIR, One Rate'', the scheme with two virtual users and
superposition coding in \eqref{eq308_6} is denoted as ``CSIR, Two Rates'', `CSIR,
Infinite Rates'' represents the approach for infinitely many users and
the achievable rate (\ref{eq308_5}), and ``CSIR, Capacity'' denotes
the capacity with receiver-only CSI in (\ref{eq302_1}). Also, the
capacity with both transmitter and receiver CSI obtained by
waterfilling \cite{goldsmith97} is shown as ``CSIRT, Waterfilling''.
We can see that the CSIR and CSIRT capacities are the same in
high SNR, indicating that transmitter CSI is not necessary in this
case.  We also observe  that the rate loss of single rate fixed transmission to the  capacity
increases as SNR increases.
For two virtual users, the achievable rate improves at high SNR,
however, still with an increasing rate loss to capacity. In contrast,
the infinite virtual user case exhibits a constant rate loss to
capacity.

\begin{figure}\centering
  \includegraphics[scale=0.45]{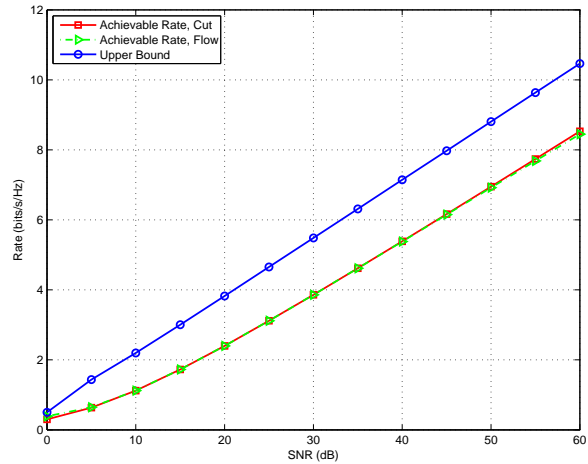}\\
  \vspace{-3mm}
  \caption{\hspace{-0.5mm}Comparison\hspace{-0.25mm} of \hspace{-0.25mm}rates \hspace{-0.5mm}for a \hspace{-0.1mm}diamond \hspace{-0.5mm}network\hspace{-0.5mm} by
    maximizing (\ref{eq403}) directly and by using the proposed
    distributed algorithm and the upper bound
    (\ref{eq410}).}\vspace{-6mm}\label{f6}
\end{figure}

We now consider a diamond network with a single source node $s$, a
single destination node $d$, and two relay nodes $r_1$ and $r_2$.
There are no direct links between source and destination node and
between the relay nodes, resp., and source and relays transmit in orthogonal
channels. Fig.~\ref{f6} displays the achievable rate and the upper
bound (\ref{eq410}) for different $P$.
As the network has only four cuts, we can compute the
achievable rate for fixed rate transmission by using the cut-set
formulation \eqref{eq403} directly.  We also include the achievable
rate using the flow-based formulation (\ref{eq435}) optimized via the
distributed algorithm in Section \ref{sect43}.
We can see that the flow-based formulation achieves the same rate as
the cut-set formulation. The rate gap between the achievable rate and
the upper bound increases as $P$ increases but diminishes relatively
to the total rate.

\vspace*{-1ex}
\section{Conclusion}\label{sect6}\vspace*{-0.7ex}
We have analyzed the achievable average rate in interference-free
networks along with the optimal rate allocation on network and
physical layers. For the point-to-point fading channel, we presented a
superposition coding approach based on multiple virtual
users to improve the throughput of fixed rate transmission when only
receiver CSI is available. For networks, a distributed cross-layer
optimization algorithm was proposed to obtain the maximum achievable
rate using a fixed rate transmission scheme on the physical layer.
It was shown that the proposed layered approach asymptotically
achieves almost all of the ergodic network capacity  in high SNR.

\vspace*{-0.5ex}
\section*{Acknowledgment}\vspace*{-0.5ex}
This material is partly funded by subcontract \#069153 issued by BAE
Systems National Security Solutions, Inc. and supported by the Defense
Advanced Research Projects Agency (DARPA) and the Space and Naval
Warfare System Center (SPAWARSYSCEN), San Diego under Contract No.
N66001-08-C-2013, by NSF grant CCF08-30666, and by Caltech's Lee Center for Advanced Networking.

\vspace{-0.2ex}
\bibliographystyle{IEEEtran} 
\bibliography{IEEEabrv,ref}

\end{document}